\documentclass[pre,
 reprint,
 amsmath,amssymb,
 aps,twocolumn, showkeys,
]{revtex4}
\usepackage{wasysym}
\usepackage{graphicx}
\usepackage{dcolumn}
\usepackage{bm}
\usepackage{epsfig,subfigure,float,pseudocode,multirow,footmisc,rotating}
\usepackage[nottoc, notlof, notlot]{tocbibind}
\usepackage{mathtools}
\usepackage{hyperref}
\usepackage{color}

\begin{document}

\title{Collective departures in zebrafish: profiling the initiators}

\author{Bertrand Collignon}
 \email{bertrand.collignon@univ-paris-diderot.fr}
\affiliation{
 Univ Paris Diderot, Sorbonne Paris Cit\'e\\
 LIED, UMR 8236, 75013, Paris, France
}

\author{Axel S\'eguret}
\affiliation{
 Univ Paris Diderot, Sorbonne Paris Cit\'e\\
 LIED, UMR 8236, 75013, Paris, France
}

\author{Yohann Chemtob}
\affiliation{
 Univ Paris Diderot, Sorbonne Paris Cit\'e\\
 LIED, UMR 8236, 75013, Paris, France
}

\author{Leo Cazenille}
\affiliation{
 Univ Paris Diderot, Sorbonne Paris Cit\'e\\
 LIED, UMR 8236, 75013, Paris, France
}

\author{Jos\'e Halloy}
\email{jose.halloy@univ-paris-diderot.fr}
\affiliation{
 Univ Paris Diderot, Sorbonne Paris Cit\'e\\
 LIED, UMR 8236, 75013, Paris, France
}

\begin{abstract}

In social animals, morphological and behavioural traits may give a stronger influence on the collective decisions to some individuals, even in groups assumed to be leaderless as fish shoals. Here, we study and characterize the leadership of collective movements in shoals of zebrafish \textit{Danio rerio}. We observe groups of 2, 3, 5, 7 and 10 zebrafish swimming in a two resting sites arena during one hour and quantify the number of collective departures initiated by each fish. While all fish attempt to initiate at least one departure, some individuals lead more departures than others in most groups. By measuring the number of attempts made by each fish, we demonstrate that they have actually the same success rate to lead the group out of a resting site after an attempt but that this success rate decreases for larger groups. Then, we show that the succession of initiators is not temporally organised as the probability to lead a departure do not depend on the previous status of the leader. Finally, we highlight that the intra-group ranking of a fish for the initiative is correlated to its intra-group ranking for the average speed with mobile individuals more prone to lead the shoal.

\end{abstract}

\keywords{Collective behaviour --- \textit{Danio rerio} --- Leadership --- Initiation}

\maketitle

\section{Introduction}

Collective departure is a decision-making faced by all social species that travel in groups. In this process, an individual generally initiates the movement of the group out of a residence site or towards a new direction. The identity and motivation of this initiator vary widely according to the social organisation of the considered species \cite{PetitAndBon.2010, Kingetal.2009}. In hierarchical societies, the leadership may be taken on by a unique or a subset of individuals that monopolise the decisions. These individuals can be older \cite{SueurAndPetit2008} or of a specific sex \cite{Rasa1983}. This results in a consistent leadership over time generally observed in stable and closed groups and linked to a dominant position in the group \cite{Petersonetal.2002, SueurAndPetit2008}. In societies that do not identify a specific individual as the group leader, initiations may be performed by any member of the group without consistency over time. In these cases, the initiators may be temporarily more motivated due to their physiological state \cite{Fischhoffetal.2007, Sueuretal.2010, Randsetal.2003}, level of information \cite{couzin2005effective, CollignonAndDetrain2010, Pillotetal.2010} or position in the group \cite{BumannAndKrause.1993, Lecaetal.2003}. This mechanism is often present in social species that live in open groups with no consistent membership like bird flocks or fish schools. Thus, although each individual can initiate collective movement in these more egalitarian societies, some characteristics may enhance the probability of some members to take the leadership more often than others.

In fish, that are generally characterized as leaderless groups \cite{Krauseetal.2000}, collective movements are mainly driven by the individuals located at the front of the shoal \cite{BumannAndKrause.1993}. Several motivations might prompt a fish to occupy these leading positions. Starved fish that have temporary higher nutritional needs are observed at the front positions of the shoal \cite{Krause.1993a} associated with a higher rate of prey capture and food intake \cite{Krauseetal.1992, Krause.1993b}. In this case the preference for leading positions dissipates once the fish are fed \cite{Krause.1993a}. Similarly, individuals that know the location of a potential food source can lead a group of naive fish towards foraging patches either by initiating departures \cite{Reebs.2000} or favoring a particular swimming direction \cite{couzin2005effective, couzin2011uninformed}. The success of this steering has been shown to be related to the size of the guiding individuals in golden shiners, larger individuals being more often followed than smaller ones \cite{Reebs.2001}. In this case, the propensity of some fish to take the lead is related to an information that can be gained by other fish or that can become irrelevant, resulting in an ephemeral leadership by some group members. Finally, the initiation of collective departures has been related to the personality of the fish (mainly \textit{bold} versus \textit{shy}) by several studies. Indeed, while front positions are linked with higher food intake, they are also more exposed to attacks by ambushed predators \cite{Krauseetal.1998b}. Faced with this trade-off, bolder individuals are more prone to exit a shelter and search for food than shyer fish that will mostly follow them rather than initiate a departure \cite{Wardetal.2004}. This asymmetry can be reinforced by the social composition of the group with shy individuals enhancing leadership by bold ones \cite{Harcourtetal.2009a}. In addition, bolder individuals show a lower behavioural plasticity than shyer ones, even when rewarded after following a partner rather than taking the lead \cite{Nakayamaetal.2013, Jollesetal.2014}. Thus, the leadership is more consistent over time in this case, even in a non-hierarchical species.

While the literature provides evidences for morphological and behavioural traits that lead some fish to become initiator more than others, the impact of this heterogeneous distribution of initiative on the collective dynamics of the group remains unclear. Indeed, most of the works rely on a preliminary binary classification of the individuals (bold or shy) that are then observed only in pairs with both fish being physically separated in two adjoined tanks or on the observation of groups during a very short period of time due to tracking limitations preventing the identification of the fish during successive collective departures. However, the recent development of tracking techniques based on the individual recognition of specific patterns associated with each fish \cite{perez2014idtracker} allows us to overcome these limitations and to individually follow fish in larger groups and for longer time periods.

In this context, we studied the distribution of the leadership in the zebrafish \textit{Danio rerio}. In its natural environment, \textit{Danio rerio} is a gregarious species that lives in small groups in shallow freshwaters \cite{Spenceetal.2008, Engeszeretal.2007, McClureetal.2006}. It is a well known model animal in genetics and neuroscience \cite{NortonAndBally-Cuif2010} but also in ethology. Studies have shown that a continuum from shy (less risk-prone) to bold (risk takers) individuals can be observed in zebrafish shoals. This inter-individual variability is correlated with the social status of the fish, bold individuals being more often aggressive and having potentially a higher reproductive success \cite{Dahlbometal.2011, AriyomoAndWatt.2012}. At the collective level, the shoaling behaviour of zebrafish is already observed in larvae and shoaling preferences appear at the juvenile stage \cite{Engeszeretal.2007b}. In adults, zebrafish periodically oscillate from loosely connected groups to dense aggregates \cite{MillerAndGerlai.2008} and regularly transit from unstructured shoals to polarised schools (and inversely). During the school phases, the zebrafish show a larger inter-individual distances and swim at a higher speed \cite{MillerAndGerlai.2012}. Thus, groups of zebrafish show a succession of mobile and static phases. Our goal is to study the initiation of such repeated short-term collective movements and the presence of initiators during successive collective departures for different group sizes. To do so, we observe groups of 2, 3, 5, 7 and 10 zebrafish swimming in an experimental arena consisting in two rooms connected by a corridor. This new experimental set-up composed of two connected patches allows us to observe a large number of collective departures for long duration experiments (1 hour) without human intervention. We expect the zebrafish to collectively rest in one of the two rooms and regularly transit towards the other one. Our aim is to measure the number of collective departures initiated by each fish and to put in relation their propensity to lead departures with the number of attempts that they made as well as their level of activity.


\section{Methods}

\subsection{Ethic statement}
The experiments reported in this study were performed in accordance with the recommendations and guidelines of the Buffon Ethical Committee (registered to the French National Ethical Committee for Animal Experiments \#40) after submission to the state ethical board for animal experiments.

\subsection{Animals and housing}

The fish were reared in housing facilities ZebTEC and fed two times a day (Special Diets Services SDS-400 Scientific Fish Food). We kept the fish under laboratory conditions, $27\,^{\circ}{\rm C}$, 500$\mu$S salinity with a 10:14 day:night light cycle. The water pH was maintained at 7 and the nitrite concentration (NO$^{2-}$) was below 0.3 mg/l. All zebrafish observed in this study were 6-12 months old at the time of the experiments.

\subsection{Experimental setup}

\begin{figure}[ht]
\centering
\includegraphics[width=.5\textwidth]{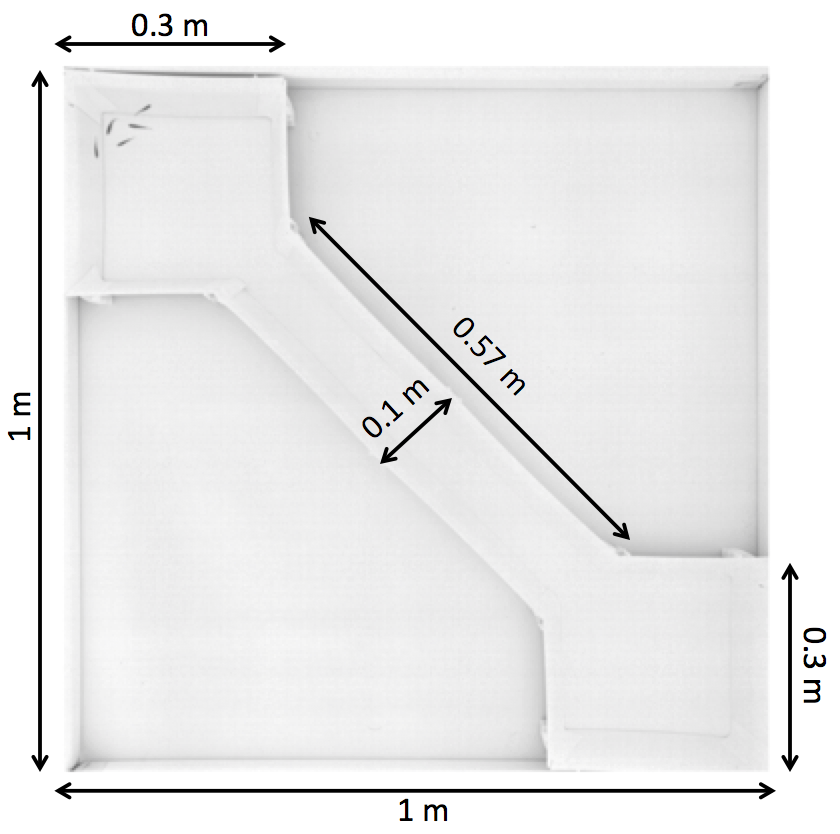}
\caption{Experimental arena consisting of two square rooms (30 cm x 30 cm) connected by a corridor (57 cm x 10 cm) placed in a 100 cm x 100 cm tank. Twelve groups of 2, 3, 5, 7 and 10 zebrafish were observed swimming freely during trials of 1 hour to study the collective departures of the fish from one room to the other.}
\label{fig:setup}
\end{figure}

We observed groups of zebrafish swimming in an arena consisting of two square rooms connected by a corridor starting at one corner of each room placed in a 100 cm x 100 cm x 30 cm experimental tank (Fig.~\ref{fig:setup}). The walls of the arena were made of white opaque PMMA. The water depth was kept at 6 cm during the experiments in order to keep the fish in nearly 2D to facilitate their tracking. One lamp (400W) was placed on the floor at each edge of the tank which is 60 cm above the floor to provide indirect lightning. The whole setup was confined behind white sheets to isolate experiments and homogenise luminosity. A high resolution camera was mounted 1.60m above the water surface to record the experiments at a resolution of 2048 x 2048 pixels and at 15 frames per second.

\subsection{Experimental procedure}

We observed 12 groups of two, three, five, seven and ten adult laboratory wild-type zebrafish (\textit{Danio rerio} AB strain) during one hour for a total of 60 experiments. Before the trials, the fish were placed with a hand net in a cylindrical arena (20 cm diameter) in one of the two rooms. Following a 5 minutes acclimatisation period, the camera started recording and the fish were released and able to swim in the experimental arena. After one hour, the fish were caught by a hand net and replaced in the rearing facilities.

\subsection{Data analysis}

The videos were analyzed offline by the idTracker software \cite{perez2014idtracker}. This multi-tracking software extracts specific characteristics of each individual and uses them to identify each fish without tagging throughout the video. This method avoids error propagation and is able to successfully solves crossing, superposition and occlusion problems. However, the tracking system failed to track correctly one experiment with two fish, one experiment with five fish and two experiments with ten fish (some sections of 5 to 10 seconds were missing on the trajectories of one or two fish). Therefore, these four experiments were excluded from our analysis. For all other experiments, we obtained the coordinates $P(x, y, t)$ of all fish at each time step $\Delta~t = 1/15s$. With these coordinates, we built the trajectories of each fish and computed their position in the arena and their instantaneous speed $v_t$ computed as the distance between $P(x,y,t-1)$ and $P(x,y,t+1)$ divided by two time steps.


\section{Results}

\begin{figure*}[ht]
\centering
\includegraphics[width=\textwidth]{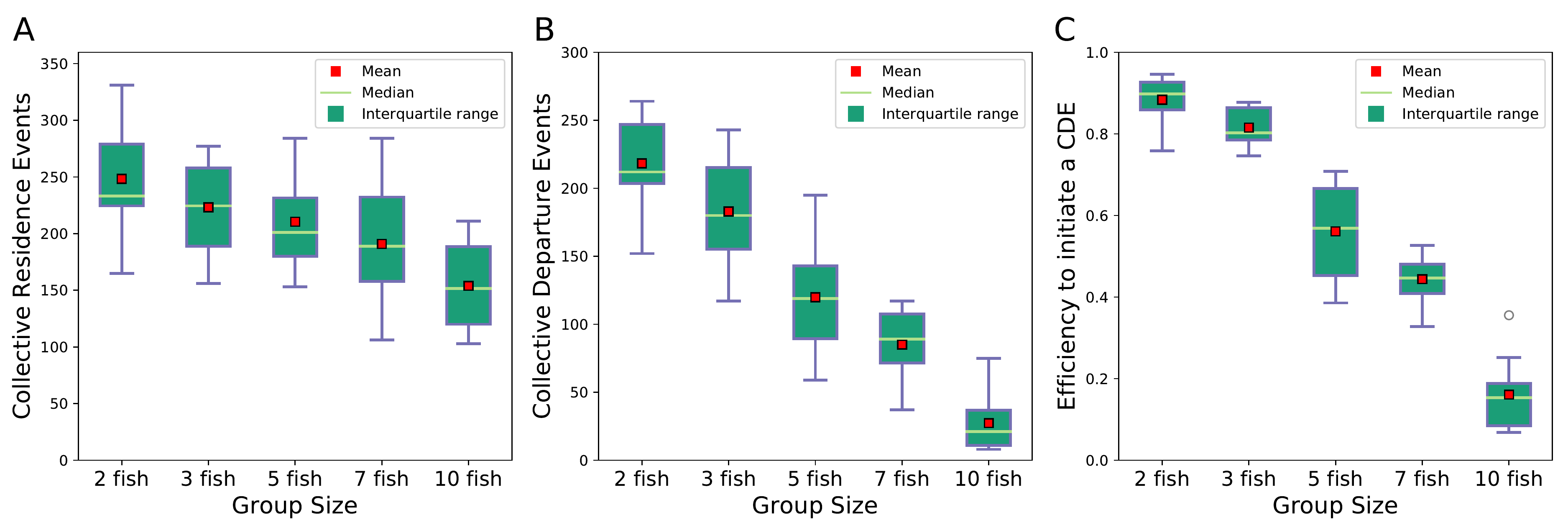}
\caption{(A) Number of collective residence events (CRE) and (B) collective departure events (CDE) for the 11 groups of two, 12 groups of three, 11 groups of five, 12 groups of seven and 10 groups of 10 zebrafish observed during one hour. Collective residence events are defined as the whole group resting in one of the two rooms and collective departures events are defined as the whole group leaving one of the resting sites. (C) Efficiency of the first leaver to trigger a collective departure of all fish computed as the proportion of CRE that were followed by a CDE.}
\label{fig:ndepartures}
\end{figure*}

In all experiments, the fish regularly aggregated in the rooms and regularly transited from one to the other. First, we quantified for all replicates the total number of collective residence events (CRE) defined as the whole group resting in one of the two rooms. The number of CRE decreases when the size of the groups increases with an average number of $248 \pm 42$ CRE for two fish to $154 \pm 36$ CRE for groups of ten fish (Fig.~\ref{fig:ndepartures}A). Then, we counted the total number of collective departures events (CDE) defined as the whole group leaving one of the resting sites for the corridor towards the other one. The number of CDE also decreases but with a stronger difference between the groups from an average number of $218 \pm 32$ for two zebrafish to $27 \pm 20$ CDE for ten zebrafish (Fig.~\ref{fig:ndepartures}B). Therefore, 88\% of the CRE were followed by a collective departure in dyads on average while only 17.5\% were in groups of ten fish (Fig.~\ref{fig:ndepartures}C). Thus, larger groups were more likely to split into subgroups during departures while small groups of fish remained cohesive most of the time.

Thanks to the individual tracking of the fish, we determined the identity of the first fish that left a room for all the collective departures. For all groups, we computed the proportion of collective departures initiated by each fish and ranked the group members according to this proportion of led departure. To characterise the distribution of the leadership among the group members, we compared these experimental distributions with two theoretical ones. On the one hand, we simulated a situation where all fish have the same probability ($1/n_{fish}$) to initiate a departure. On the other hand, we simulated a despotic configuration with a fish that has a 0.9 probability to initiate a collective movement while the others have only a $0.1/(n_{fish}-1)$ chance to start a departure. The experimental data lays between these two extreme scenarios (Fig.~\ref{fig:rankedinitiation} for groups of 5 fish). In groups of five fish, the 1$^{st}$ ranked fish initiated 45\% of the collective departures on average. This value is largely below the 90\% observed in the despotic situation but also higher than the 25\% of the uniform repartition. We observed similar results for groups of 2, 3, 7 and 10 fish (see supplementary figure 1) and compared the distributions of the initiations in each group with a homogeneous distribution by a $\chi^2$ test of goodness of fit. Among the 56 groups, only four groups of 2 fish, one group of 3 fish and 4 groups of 10 fish did not significantly differ from the equal repartition of the leadership (see supplementary table 1 for details). These results highlight a heterogeneous distribution of leadership among group members (although not despotic) in the majority of the groups with some fish having a higher tendency to start a departure than others.

\begin{figure}[ht]
\centering
\includegraphics[width=.5\textwidth]{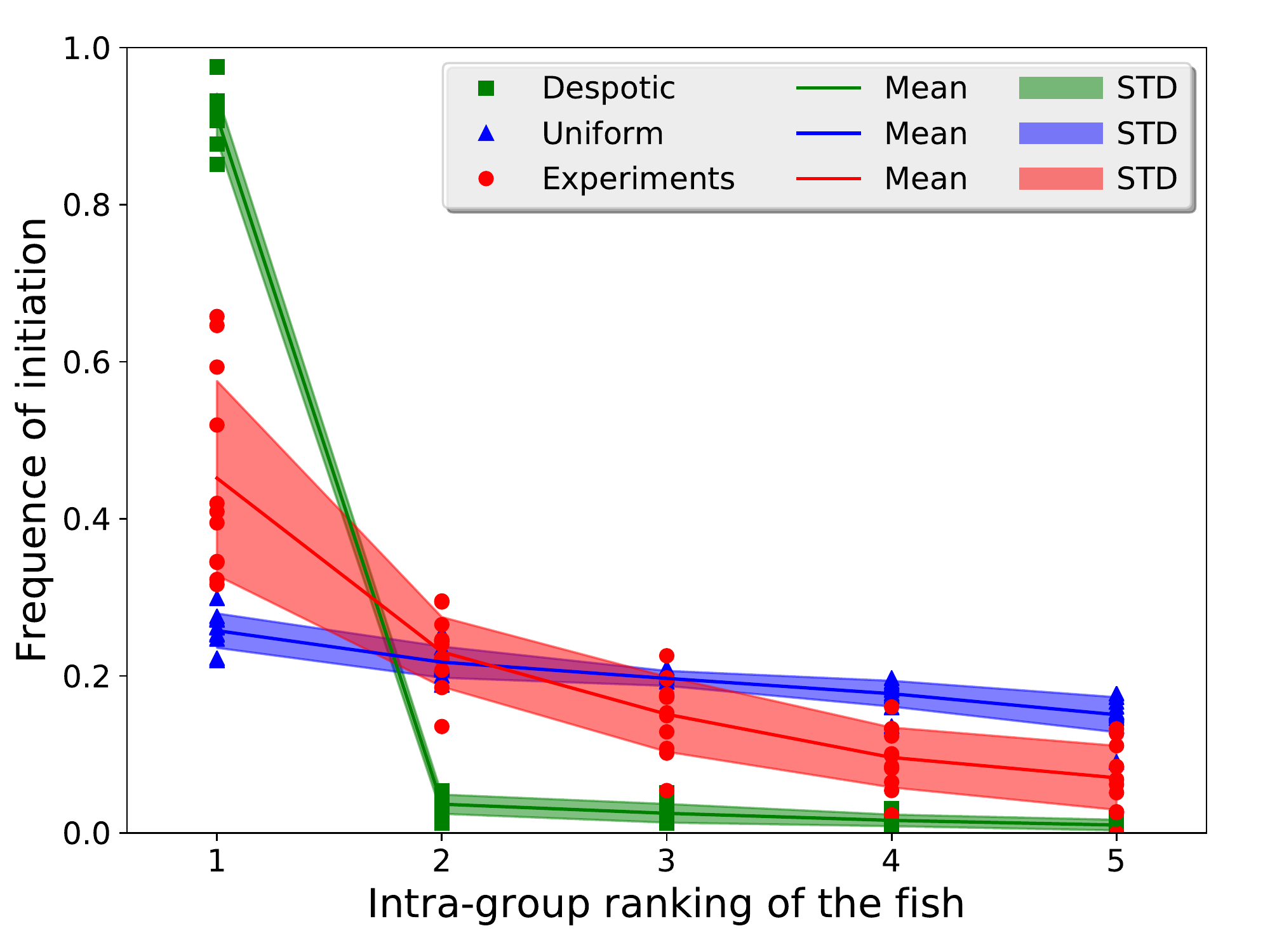}
\caption{Frequency of initiation according to the intra-group ranking of the fish in groups of 5 individuals. For each group, we computed the proportion of departure initiated by each fish and ranked them according to this frequency. The experimental results (in red) are compared with simulated distributions (uniform in blue and despotic in green). The uniform distribution assumes that the fish have the same probability $p = 0.2$ to initiate a collective departure while the despotic distribution assumes that one fish has a probability $p = 0.9$ and the other a probability $p = 0.025$ to initiate a departure.}
\label{fig:rankedinitiation}
\end{figure}

\begin{figure}[ht]
\centering
\includegraphics[width=.5\textwidth]{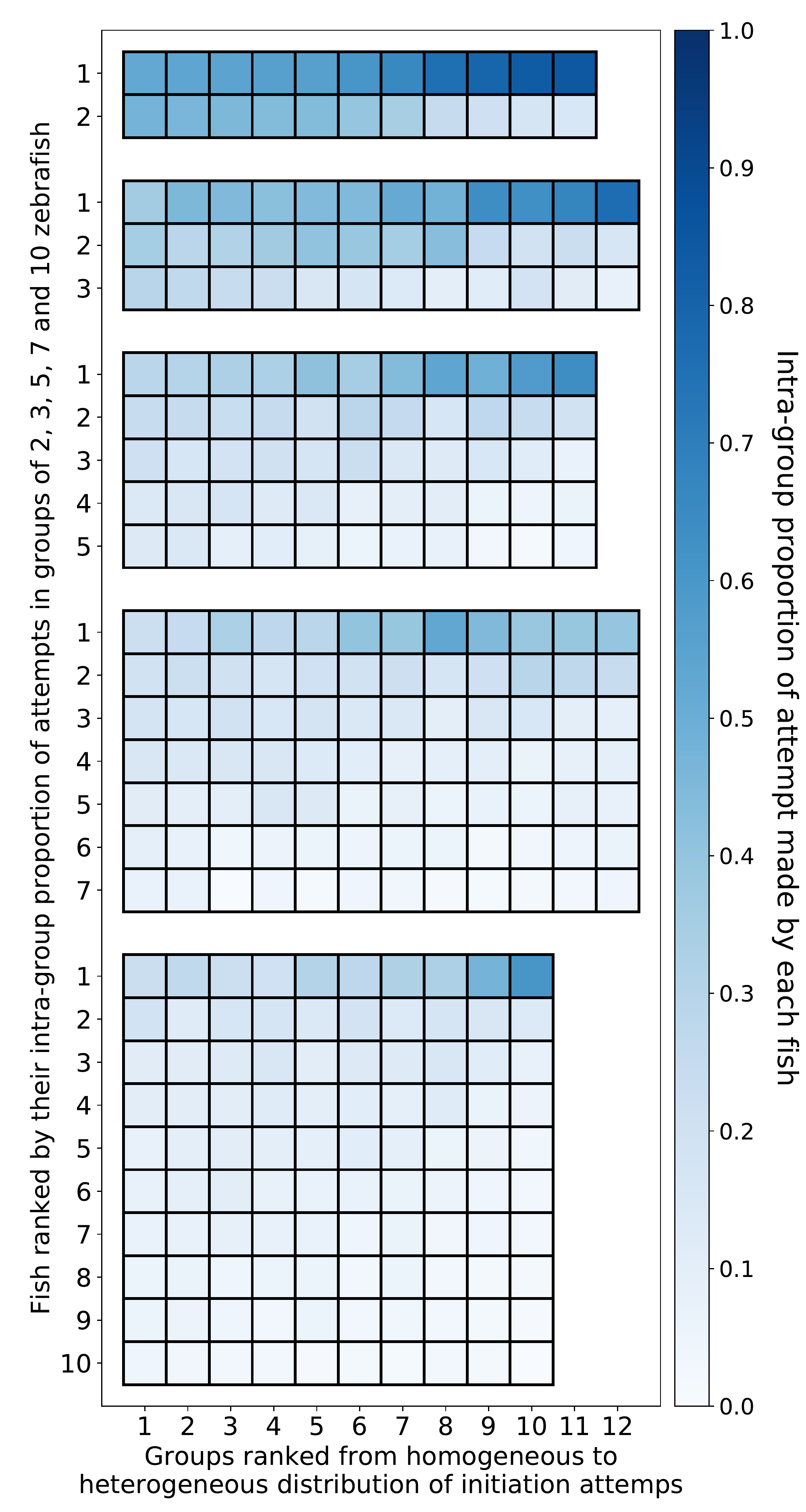}
\caption{Proportion of initiation attempt made by each fish for group of 2, 3, 5, 7 and 10 fish. For each population size, the groups are ranked from the most homogeneous distribution of attempts (left) to the most heterogeneous distribution (right). In each group, the fish are ranked from the highest proportion of attempt (top) to the lowest proportion of attempt (down).}
\label{fig:prop_attempt}
\end{figure}

To determine whether this distributed leadership was related to a different success rate or to a different number of initiation attempts, we measured the number of times that each fish was the first to exit a resting site independently of its success to be followed by the other group members (defined as an \textit{attempt}). For each group, we compared the distribution of the total number of attempts among the group members with a theoretical homogeneous distribution ($\chi^2$ test of goodness of fit with a homogeneous distribution, see supplementary table 2 for details). For two fish, 4 dyads out of 11 did not significantly differ from the homogeneous distribution. In groups of three fish, the hypothesis of homogeneous distribution was not rejected in only one trio. Finally, all groups of 5, 7 and 10 fish significantly differ from the equal distribution of the number of attempts. In addition, we computed the proportion of attempts made by each fish and ranked them according to their score. We also ranked each group according to the level of deviation from the homogeneous distribution (measured by the $p$-value of the $\chi^2$ test). These rankings show the presence of a continuum from an more egalitarian to a more despotic distribution for all group sizes (Fig.~\ref{fig:prop_attempt}). Thus, like the distribution of initiations, the attempts are generally heterogeneously distributed among the members of a group.

\begin{figure}[ht]
\centering
\includegraphics[width=.45\textwidth]{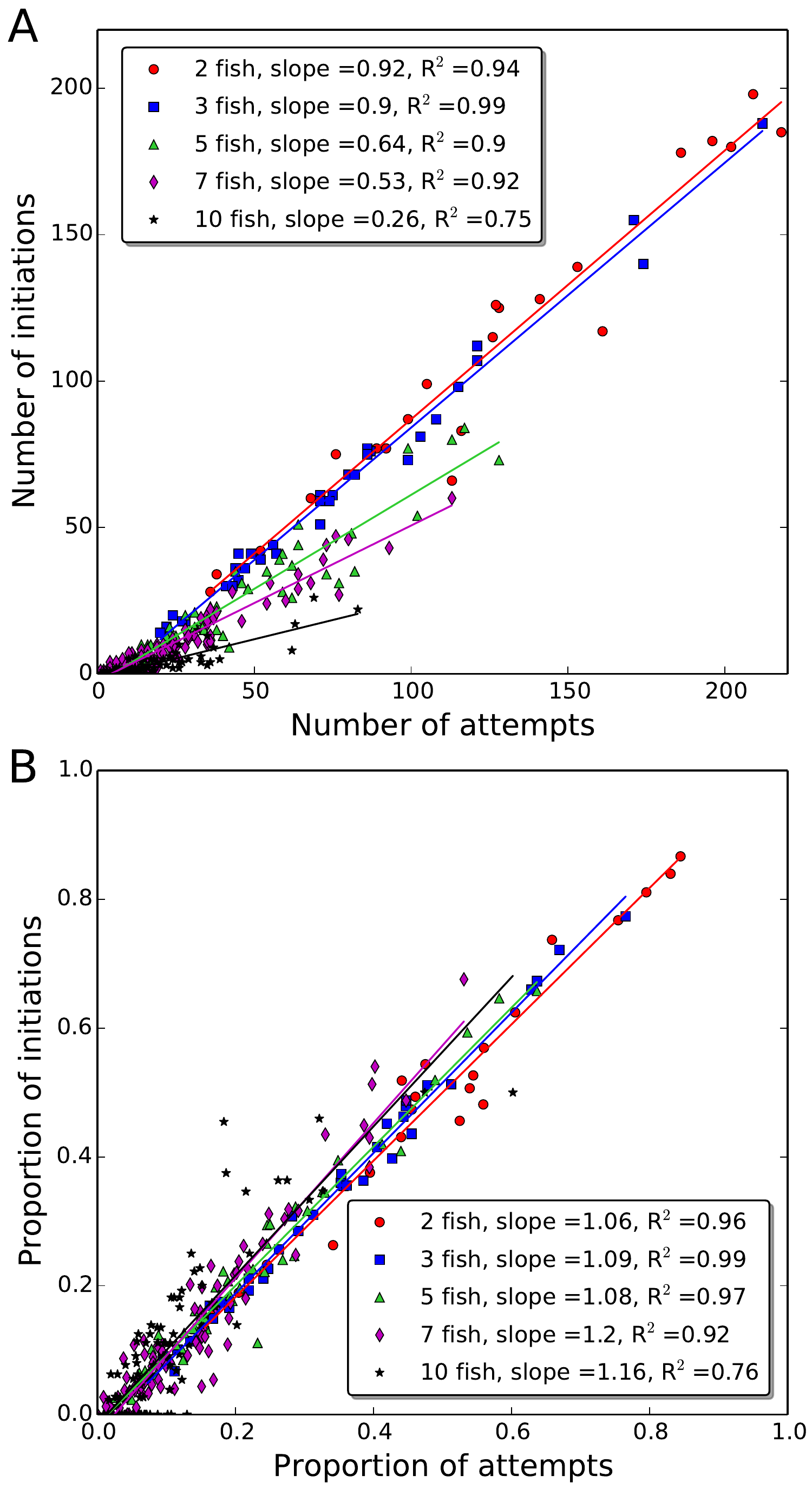}
\caption{(A) Total number of collective departures initiated as a function of the total number of attempts for each fish. The number of initiations is directly proportional to the number of attempts but the success rate of the initiations decreases for larger group sizes. (B) Proportion of attempts made by the fish in relation to the proportion of departures initiated. For each group size, the success rate is identical for all fish in the shoal.}
\label{fig:attempts}
\end{figure}

Therefore, we analyzed the potential correlation between the number of initiations and the number of attempts made by each fish. A linear regression shows that the number of initiations is linearly correlated to the number of attempts performed by the fish and that the coefficient of this correlation depends on the group size (Fig.~\ref{fig:attempts}A). For groups of two fish, 92\% of the attempts made by an individual resulted in a collective departure of the dyad. In accordance with the results observed at the group level (Fig.~\ref{fig:ndepartures}C), the success rate for each fish decreases when the population increases: 90\% for 3 fish, 64\% for 5 fish, 53\% for 7 fish and 26\% for 10 fish. While the probability of group splitting increases with the group size, the linear relation between the numbers of attempts and initiations highlights that this success rate to initiate a collective departure is the same for all fish. This conclusion is confirmed by the intra-group proportion of initiation led that is equal to the proportion of attempts performed by each fish (Fig.~\ref{fig:attempts}B). Thus, the larger number of led departures by some fish is not related to a higher influence on other group members or a better success rate but on a higher tendency to exit the resting sites.

\begin{figure}[ht]
\centering
\includegraphics[width=.45\textwidth]{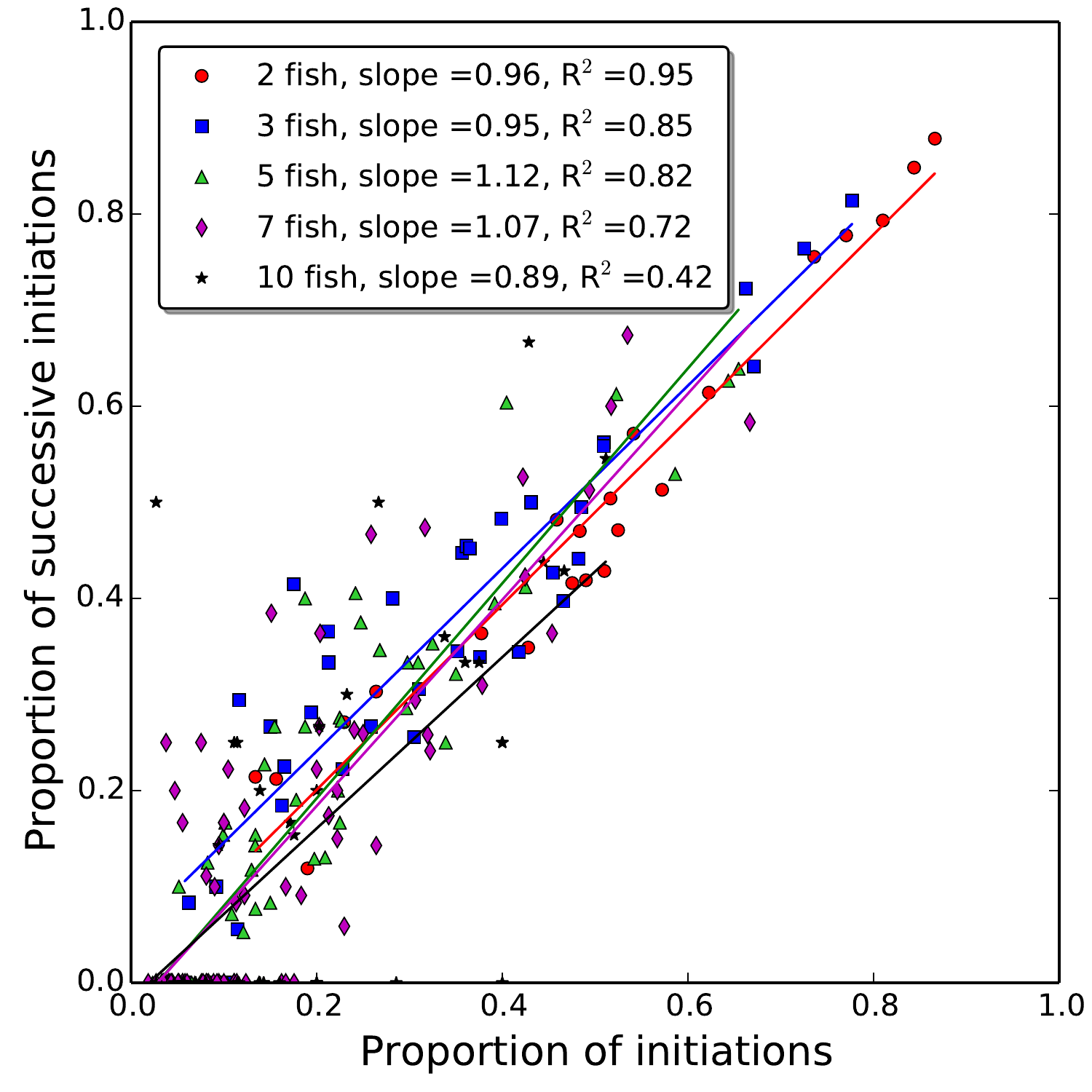}
\caption{Proportion of successive initiations as a function of the proportion of initiated collective departures for 2, 3, 5, 7 and 10 fish. The slopes of the line $\approx1.0$ highlight that the leadership events of each fish were homogeneously distributed along the experiment without temporal segregations of the initiators. The initiation of a departure does not increase the probability to initiate the following one.}
\label{fig:samelead}
\end{figure}

Next, we studied the temporal distribution of the leading events to highlight a potential temporal organisation of the leaders over successive departures. To do so, we computed the probability to observe a fish performing two successive initiations and compared this probability to the proportion of departure that the fish has led. A temporal segregation of the initiators would results in a high probability of successive initiations compared to the proportion of led departure while homogeneously distributed initiations would give similar probabilities of successive initiations and led departure. For all group sizes, we observed a linear and direct relation between the two proportions (Fig.~\ref{fig:samelead}). Thus, the probability of a fish to initiate a CDE was not dependent on its status of initiator or follower during the previous CDE.

\begin{figure*}[ht]
\centering
\includegraphics[width=\textwidth]{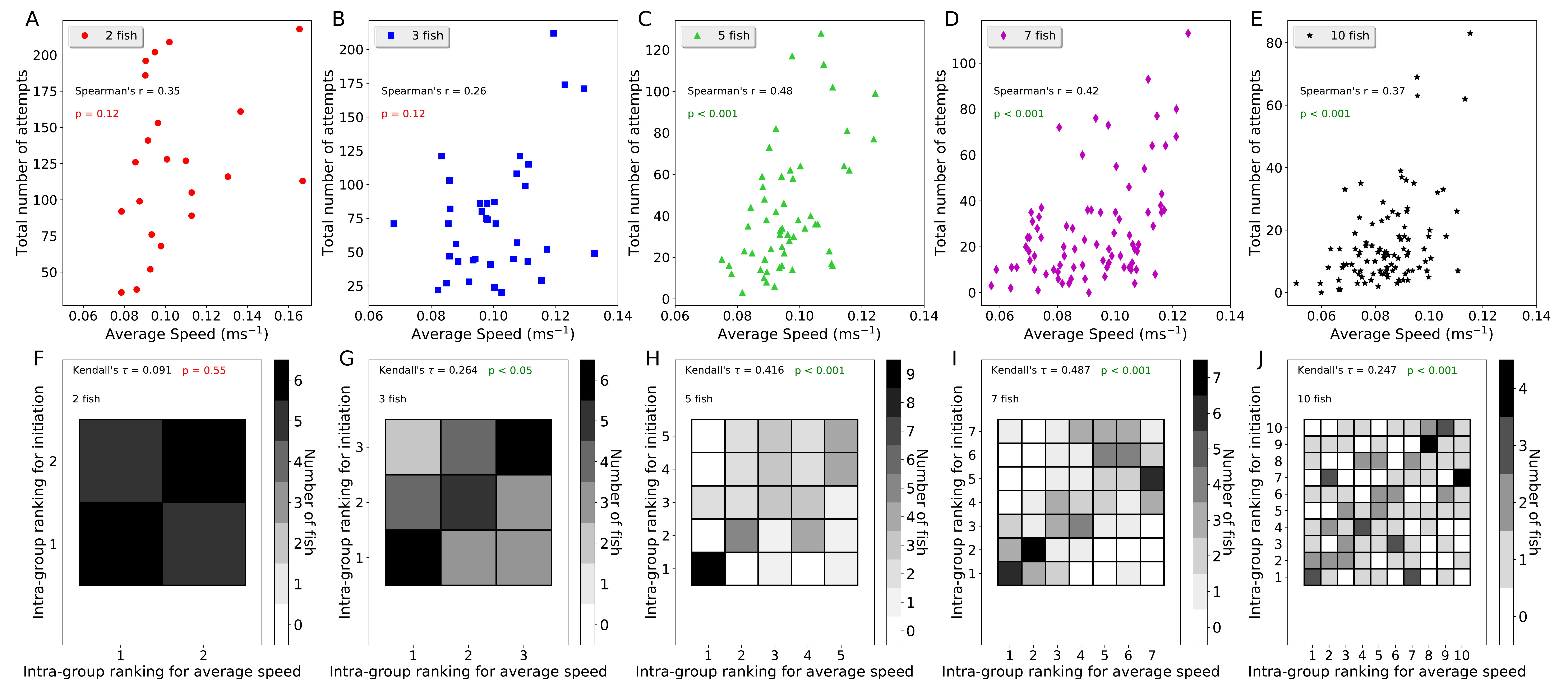}
\caption{(A-E) Proportion of attempt made by each fish according to its average speed. For small group sizes (2 or 3 fish), the absolute value of the mean linear speed is not a good predictor of the number of attempts performed by a fish (Sperman's correlation) but for larger group size (5, 7 and 10 fish), the linear speed of a fish is statistically correlated with the number of attempts. (F-J) Distribution of the fish according to their intra-group ranking of the number of collective departures that they initiated and their intra-group ranking for the average speed. By taking into account their ranking inside the group, the relationship is statistically significant as soon as the groups is formed by at least three fish.}
\label{fig:speedlead}
\end{figure*}

Finally, we looked at a potential link between the motion characteristics of the fish and the number of attempts that they have made. In particular, we measured the average linear speed of all individuals as an indicator of their motility. There is a positive correlation between the average speed of the fish and the number of attempts that they performed. However, this correlation is only significant for groups of 5, 7 and 10 fish (Fig.~\ref{fig:speedlead}A-E, Spearman's correlation). As the number of attempts made by a fish depends on the motivation of its groupmates (a potential \textit{very motivated} initiator could be hidden by a \textit{super motivated} initiator), we also compared the intra-group ranking of the fish for the number of initiations with their intra-group ranking for the linear speed. We used the Kendall's $\tau$ coefficient to measure the association between the two rankings. The intra-group ranking for the initiation is positively correlated to the intra-group ranking for the linear speed (Fig.~\ref{fig:speedlead}F-J) for groups of 3, 5, 7 and 10 fish. So, except for dyads, the fish with the highest average speed of its group is more likely to also be the fish that has started the largest number of departures. Thus, the initiation of collective movements is related to the motility of the fish.

\section{Discussion}

The initiation of collective movements in fish is often reported as a distributed process in which each fish can potentially lead a departure. This distribution of the leadership is particularly suited for large schools of hundreds or thousands of fish that have to detect and avoid attacks from predators coming potentially from any direction \cite{Krauseetal.2000}. Thus, the first fish to spot a predator can start an escaping manoeuvre that will be propagated from neighbour to neighbour in the whole school. However, numerous fish species live in shoals that do not exceed one or a few dozens of individuals. In such smaller groups, some individuals might have a stronger influence than others on the collective movement of the shoal. To investigate this question, we studied the initiation of collective movement in groups of zebrafish \textit{Danio rerio} swimming in an environment composed by two spots in a non-stress situation. In this context, we showed that the initiation of collective departures is a distributed process among the group members. However, the role of initiator is not homogeneously distributed, with some individuals leading more departures than others. By measuring the number of attempts made by each fish, we highlighted that this heterogeneous distribution was not the result of a higher success rate of some individuals that could have a higher tendency to be followed. On the contrary, the success of the fish to trigger a collective departure was linearly correlated to their number of attempts. A similar result was also observed in other fish species like Damselfish in which collective departures from one spot to another was mainly led by fish that performed a higher number of attempts \cite{Wardetal.2013}. In addition, we showed that the initiation process was not temporarily organised with a fish leading the group during a particular time period before being relayed by another fish, but was distributed during the whole experimental time among the group members.

While the linear relation between the number of attempts and initiations was observed for all group sizes, the success rate of the attempts drops from $\approx 90\%$ in dyads to only $\approx 20\%$ in groups of ten fish. The majority of attempts led to a temporary fission of the shoal into subgroups for this larger population size. In our experimental setup, the subgroups always reassembled after a short period of time. However, in natural conditions where the fish are not restrained to a small environment, those splitting events could lead to a consistent fission of the group. Indeed, zebrafish form shoals of a few to a dozen of individuals in their habitat \cite{Parichy2015}. The size of the shoals observed in nature can be driven by a trade-off between the advantages (e.g. detection of predators and potential food sources) and disadvantages (e.g. larger groups are more easily spotted by predators, increased inter-individual competition for food) of being in groups. According to our results, the intermediate shoal sizes observed in zebrafish could be maintained by a strong cohesion in small groups but a loose organisation of larger ones making them more prone to splitting.

Our results also highlighted that the motility of the fish is a predictor of its tendency to initiate collective departures. Indeed, except for duos, the intra-group ranking of a fish for the average speed was correlated to its intra-group ranking for the number of led departures. Therefore, the leaders of collective movements in zebrafish do not seem to occupy a particular hierarchical status in the group but are generally the most mobile individuals. A similar result was predicted by a theoretical analysis on the emergence of leadership in simulated zebrafish \cite{Zienkiewiczetal.2015}. This study showed that an informed individual moving in a specific direction is more likely to be followed by a group of naive individuals when it moves just faster than the naive group. In Damselfish \textit{Dascyllus aruanus}, the initiator of a collective movement also displays a higher level of activity than their group members before the departure \cite{Wardetal.2013}. A favored direction, a higher level of activity or a higher average speed can lead a fish to occupy the front position of the shoal more often than its group members. As the direction of the group is mainly decided by the front individuals \cite{BumannAndKrause.1993}, these inter-individual behavioural differences lead to a heterogeneously distributed leadership in the shoal. 

Finally, our results also show that the sharing of the leadership across the different groups is a continuum from a homogeneously distributed leadership to strongly asymmetrical distributions. A similar diversity was observed in groups of four zebrafish locating a food patch \cite{perez2014idtracker}: in two groups out of four, the order of arrival was consistent over successive trials while the fish in the two other groups showed a random arrival order. In our experiments, dyads showed the most egalitarian situations but also the strongest monopolization of leadership with one fish performing up to 85\% of the initiations of its group. A similar result was observed in trios with some groups sharing equally the leadership between all group members and other groups with a disproportionate number of initiations led by the same fish (up to 75\% for one group). As the group size increases, almost all groups showed a heterogeneous distribution of the leadership between the fish even if we did not observe a clear monopolization of the initiations of collective departures in these shoals. A similar effect of group size on leader-followers interaction was evidenced in minnows \cite{Partridge.1980}. In this latter study, 6 out of 9 dyads displayed a clear leader-follower relation, 2 showed an equally shared leadership and 1 was formed by fish that did not interact with each other. The author concluded that one fish leads the other in groups of two but that this behaviour is not observed for larger groups.

Stronger asymmetries are more likely to be observed in small group sizes but an unbalanced distribution is almost always present in groups of a dozen of individuals. Such outcome can be the results of sampling of a continuous distribution for an individual characteristic that influences the probability to lead the group. Indeed, as we add more individuals, there is a higher probability that at least two of them significantly differ from each other, leading to an unshared decision-making process but by the same time, the average difference between individuals tends to stabilize to a limit value. On the contrary, as only two fish are forming a dyad, there is a probability that these fish are either almost identical, resulting in a homogeneous leadership, or on the contrary strongly different, leading to a heterogeneous leadership, with a continuum of possibilities between these extrema. Incidentally, one should be cautious when subdividing the tested population into binary behavioural classes and then performing experiments with pairs of opposed individuals. Indeed, such classification may lead to a misrepresentation of the social organisation observed in free groups by forming only asymmetrical pairs. This effect was already mentioned by \cite{LeblondAndReebs.2006} that concluded that a significant relation between boldness and leadership in golden shiners only when they classified the fish into the binary \textit{leaders} or \textit{non-leaders} classes. Thus, in animal species in which the initiation of specific behaviours is related to individual characteristics rather than to a particular hierarchical position, are more likely to display a whole range of social structures from despotic to egalitarian groups without any behavioural changes but only due to sampling effects.

\phantomsection

\section*{Authors' contribution}
B.C. designed the study, performed the data analysis, ran the simulations and wrote the manuscript. A.S. and Y.C. performed the experiments and collected the experimental data. L.C. developed the video treatment and tracking system. J.H. conceived, coordinated the study and revised the manuscript. All authors gave final approval for publication.

\section*{Competing interests}
All authors declare having no competing interests.

\section*{Funding}
This work was supported by European Union Information and Communication Technologies project ASSISIbf, (Fp7-ICT-FET n. 601074). The funders had no role in study design, data collection and analysis, decision to publish, or preparation of the manuscript.


\newpage

\begin{figure*}[ht]
\centering
\includegraphics[width=\textwidth]{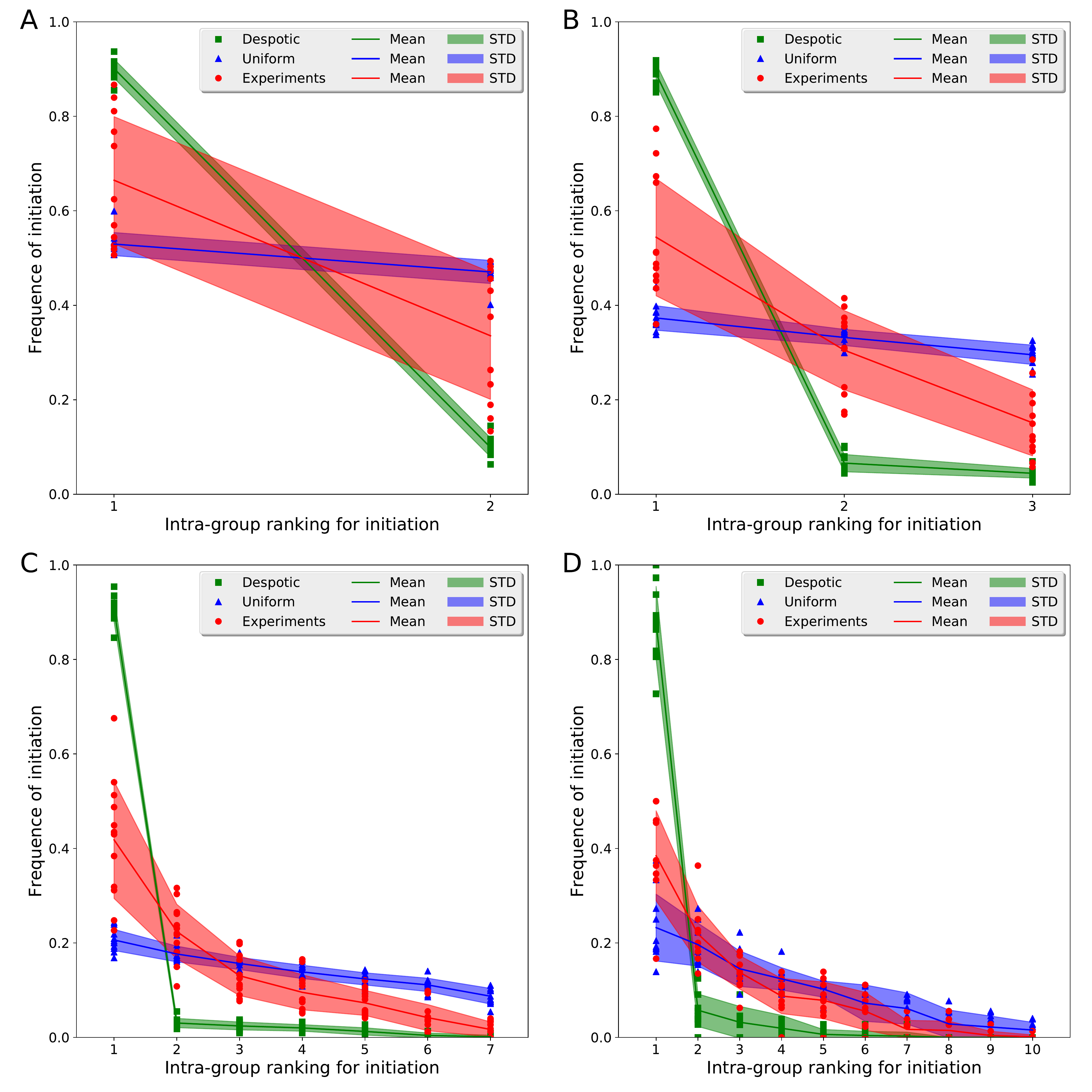}
\caption{Supplementary figure 1. Frequency of initiation according to the intra-group ranking of the fish in groups of (A) two, (B) three, (C), seven and (D) 10 individuals. For each group of \textit{n} fish, we computed the proportion of departure initiated by each fish and ranked them according to this frequency. The experimental results (in red) are compared with simulated distributions (uniform in blue and despotic in green). The uniform distribution assumes that the fish have the same probability $p = \frac{1}{n}$ to initiate a collective departure while the despotic distribution assumes that one fish has a probability $p = 0.9$ and the other a probability $p = \frac{0.1}{n-1}$ to initiate a departure.}
\label{fig:rankedinitiation}
\end{figure*}

\newpage

\begin{table}
\caption{Supplementary table 1. $p$-values of the $\chi^2$ test of goodness to fit the number of initiation made by the fish in each group with a homogeneous distribution.}
\begin{ruledtabular}
\begin{tabular}{c | c | c | c | c | c}
Group &2 fish& 3 fish & 5 fish & 7 fish & 10 fish\\
\hline
1 & 0.87 & 0.32 & 6.5 $10^{-3}$ & 3.3 $10^{-2}$ & 0.36 \\
2 & 0.56 & 4.9 $10^{-2}$ & 6.2 $10^{-3}$ & 2.6 $10^{-3}$ & 0.22 \\
3 & 0.38 & 4.1 $10^{-4}$ & 4.7 $10^{-4}$ & 3.2 $10^{-6}$ & 0.21 \\
4 & 0.23 & 1.9 $10^{-4}$ & 1.1 $10^{-5}$ & 3.3 $10^{-7}$ & 0.13 \\
5 & 4.9 $10^{-2}$ & 3.2 $10^{-7}$ & 2.0 $10^{-6}$ & 2.7 $10^{-9}$ & 1.8 $10^{-3}$\\
6 & 3.7 $10^{-4}$ & 2.1 $10^{-8}$ & 1.6 $10^{-11}$ & 9.2 $10^{-17}$ & 4.5 $10^{-4}$\\
7 & 5.8 $10^{-14}$ & 3.7 $10^{-9}$ & 9.6 $10^{-12}$ & 5.0 $10^{-18}$ & 3.2 $10^{-5}$\\
8 & 8.6 $10^{-18}$ & 3.4 $10^{-14}$ & 2.1 $10^{-15}$ & 6.1 $10^{-19}$ & 6.4 $10^{-10}$\\
9 & 2.0 $10^{-20}$ & 1.8 $10^{-19}$ & 7.9 $10^{-25}$ & 2.4 $10^{-19}$ & 1.1 $10^{-13}$\\
10 & 4.6 $10^{-23}$ & 3.6 $10^{-25}$ & 8.1 $10^{-33}$ & 3.8 $10^{-23}$ & 4.3 $10^{-17}$\\
11 & 2.2 $10^{-26}$ & 1.2 $10^{-30}$ & 3.3 $10^{-38}$ & 2.3 $10^{-24}$ & - \\
12 & - & 9.6 $10^{-48}$& - & 2.3 $10^{-33}$ & - \\
\end{tabular}
\end{ruledtabular}
\label{tab:table_init}
\end{table}

\begin{table}
\caption{Supplementary table 2. $p$-values of the $\chi^2$ test of goodness to fit the number of attempts made by the fish in each group with a homogeneous distribution.}
\begin{ruledtabular}
\begin{tabular}{c | c | c | c | c | c}
Group &2 fish& 3 fish & 5 fish & 7 fish & 10 fish\\
\hline
1 & 0.46 & 0.37 & 3.5 $10^{-4}$ & 1.8 $10^{-3}$ & 3.4 $10^{-5}$\\
2 & 0.31 & 5.3 $10^{-3}$ & 3.3 $10^{-5}$ & 2.8 $10^{-5}$ & 9.4 $10^{-7}$\\
3 & 0.14 & 2.7 $10^{-3}$ & 1.5 $10^{-5}$ & 4.6 $10^{-10}$ & 8.0 $10^{-8}$\\
4 & 0.072 & 1.5 $10^{-3}$ & 4.6 $10^{-6}$ & 8.9 $10^{-11}$ & 5.3 $10^{-10}$\\
5 & 4.5 $10^{-2}$ & 8.4 $10^{-7}$ & 2.4 $10^{-11}$ & 1.7 $10^{-17}$ & 6.7 $10^{-11}$\\
6 & 1.3 $10^{-3}$ & 5.7 $10^{-8}$ & 9.9 $10^{-19}$ &4.4 $10^{-25}$ & 1.0 $10^{-12}$\\
7 & 7.9 $10^{-9}$ & 3.6 $10^{-10}$ & 6.3 $10^{-22}$ & 8.2 $10^{-26}$ & 1.7 $10^{-23}$\\
8 & 2.4 $10^{-17}$ & 5.3 $10^{-15}$ & 5.9 $10^{-23}$ & 1.5 $10^{-26}$ & 1.4 $10^{-32}$\\
9 & 4.9 $10^{-21}$ & 6.7 $10^{-19}$ & 6.4 $10^{-34}$ & 5.8 $10^{-30}$ & 1.8 $10^{-59}$\\
10 & 4.7 $10^{-23}$ & 6.4 $10^{-24}$ & 6.9 $10^{-45}$ & 1.7 $10^{-32}$ & 6.9 $10^{-60}$\\
11 & 8.2 $10^{-26}$ & 2.2 $10^{-30}$ & 1.4 $10^{-53}$ & 4.4 $10^{-37}$ & - \\
12 & - & 5.6 $10^{-52}$& - & 2.1 $10^{-40}$ & - \\
\end{tabular}
\end{ruledtabular}
\label{tab:table1}
\end{table}

\end{document}